\documentstyle[aps,prd]{revtex}

\begin{document}
\title{Quintessence-like Dark Matter in Spiral Galaxies}
\author{F. Siddhartha Guzm\'{a}n\footnote{E-mail: siddh@fis.cinvestav.mx},
Tonatiuh Matos}
\address{Departamento de F\'{\i}sica,
Centro de Investigaci\'on y de Estudios Avanzados del IPN,
A.P. 14-740, 07000 M\'exico D.F., MEXICO.
}
\author{Dar\'{\i}o N\'u\~nez and Erandy Ram\'{\i}rez}
\address{Instituto de Ciencias Nucleares, 
Universidad Nacional Aut\'{o}noma de M\'{e}xico, 
A.P. 70-543, 04510 M\'{e}xico D.F., MEXICO.}
\date{\today}
\maketitle

\begin{abstract}
Through the geodesic analysis of a static and axially symmetric space time,
we present conditions on the state equation of an isotropic perfect fluid
$p=\omega d$, when it is considered as dark matter in spiral galaxies. The 
main conclusion is that it can be an exotic fluid ($-1<\omega <-1/3$) as it
is found for Quintessence at cosmological scale.
\end{abstract}



There is no doubt about the importance of the mystery concerning the nature
of dark matter in the Universe and in particular in galaxies. The
consequences of observations made on SNIa supernovae \cite{riess} have posed
challenges to the available theoretical machinery, and certain models
explaining such phenomena have arised which propose exotic types of matter
and therefore unusual equations of state, such as a Cosmological Constant,
Cold Dark Matter models, Dilaton Fields \cite{luis} and Quintessence \cite
{stein}. However, at the galactic level there are no models consistent with
the cosmological ones, and which give some light in the understanding of the
nature of dark matter.\newline

In order to be precise about the problem let us recall the situation of the
galactic dark matter, for which we confine ourselves to the observations
made by Rubin et al. \cite{rubin} who found that for a few sample of spiral
galaxies the interstellar gas and stars lying far away from the center (in
the equatorial plane) of the corresponding galaxy behaves in a non
Kepplerian way, but their circular velocity seems to be independent of the
radius starting from a certain distance to the galactic center, i.e. the
rotation curve profile of a spiral galaxy is flat outside a central galactic
region. It was then inferred a distribution $\sim 1/r^{2}$ of non luminous
matter (dark matter) which should contribute to the flatness of the rotation
curves. There exists certain controversy about the flatness of such curves 
\cite{apj}, but in general it is accepted that rotation curves are flat up
to the precision of the meassurementes made by the astronomers and that this
behavior is reproduced even for large samples of spiral galaxies \cite
{persic}.\newline

The most accepted scenario for a spiral galaxy reads as follows: it is an
object composed by a luminous disc whose density distributed in an
exponentially decay which conspires with a dark halo whose density is
distributed as $\sim 1/r^{2}$ \ \ \cite{peebles}. In this way it is found an
explanation for the kinetic behavior of gas and stars composing a spiral
galaxy, but how was this mixture formed and what inspired nature to conspire
in this way and not another one? If the dark matter is baryonic such as
MACHOS for instance, why does its density have a non exponential
distribution as luminous matter density does? If it is non baryonic, what is
it made of, or at least which is its equation of state? This last question
is the one that occupies ourselves in the present work.\newline

In this letter we proceed in the following way: Assume that a spiral galaxy
lies on a background axysimmetric static space time, which is characterized
by the presence of a perfect fluid with an arbitrary equation of state, i.e. 
$p=\omega d$ being $\omega $ a free function, and then we find conditions
over $\omega $ that permit flat rotation curves of test particles. Other
types of candidates to dark matter are discused in
\cite{DM-CQG,Nvo}.\newline

First of all it must be clear that our treatment is valid only in the dark
matter dominated region, i.e. where the rotation curves are flat, and we do
not consider the galactic core region.\ Observational data show that the
galaxies must be composed by almost 90\% of dark matter, distributed at the
halo, in order to explain the observed dynamics of particles in the luminous
sector of the galaxy. We can thus assume that luminous matter does not
contribute in a very important way to the total energy density of the halo
of the galaxy in the mentioned region. On the other hand, the exact symmetry
of the halo is still unknown, but it is reasonable to suppose that the halo
is symmetric with respect to the rotation axis of the galaxy, so we choose
the space time to be axial symmetric. Furthermore, the rotation of the
galaxy does not have a large effect on the motion of test particles around
it; dragging effects in the halo of the galaxy can be considered too small
to seriously affect the motion of tests particles (stars) traveling around
the galaxy. The circular velocity of stars (like the Sun) is of the order of
230 Km/s, much less than the speed of light. Hence, in the region of
interest we can suppose the space-time to be static as well.\newline

Therefore, we start by considering the background described by the following
line element:

\begin{equation}
ds^{2}=-e^{2\,\psi }dt^{2}+e^{-2\psi }[e^{2\gamma }(d\rho ^{2}+dz^{2})+\mu
^{2}d\varphi ^{2}],  \label{metric}
\end{equation}

\noindent which corresponds to an static axially symmetric space-time; the
coordinates are the usual cylindrical ones.\newline

We recall the reader that observations are made upon objects lying in the
galactic equatorial plane, thus the Lagrangian for a test particle
travelling on such slide of the space time described by (\ref{metric}) is

\begin{equation}
2{\cal {L}}=-e^{2\,\psi }\dot{t}^{2}+e^{-2\psi }[e^{2\gamma }\,\dot{\rho}
^{2}+\mu ^{2}\,\dot{\varphi}^{2}],  \label{lagrangian}
\end{equation}

\noindent where dot means derivative with respect to the proper time $\tau $
of the test particle.\ The radial geodesic motion equation is then

\begin{equation}
\dot{\rho}^{2}-e^{2(\psi -\gamma )}\,\left[ E^{2}e^{-2\psi }-L^{2}\,\frac{%
e^{2\psi }}{\mu ^{2}}-1\right] =0.  \label{geodesic}
\end{equation}

\noindent where $E$, and $L$, are constants associated with this geodesic
motion along the equatorial plane.\newline

We are interested in circular and stable motion of test particles, therefore
the following conditions must be satisfied\newline

i) $\dot{\rho}=0$, circular trajectories\newline

ii)${\frac{{\partial V(\rho )}}{{\partial \,\rho }}}=0$, extreme ones\newline

iii)${\frac{{\partial^2 V(\rho)}}{{\partial\,\rho^2}}}|_{extr}>0$, and
stable.\newline

\noindent being $V(\rho )=-e^{2(\psi -\gamma )}\,\,\left[ E^{2}e^{-2\psi
}-L^{2}\,e^{2\psi }/\mu ^{2}-1\right] $.\newline

Recalling that $E$ and $L$ are constants of motion for each circular orbit,
it is straightforward to obtain expressions for the energy $E$, angular
momentum $L$, angular velocity $\Omega =d\varphi /dt$ and the tangential
velocity $v^{(\varphi )}={e^{-2\,\psi }}\mu \,\,\Omega $ \ \ \ \ \cite
{chandra}, corresponding to a circular, stable equatorial motion:

\begin{eqnarray}
E &=&e^{\psi }\,\sqrt{{\frac{{{\frac{{\mu _{,\rho }}}{{\mu }}}-\psi _{,\rho}}%
}{{{\frac{{\mu _{,\rho }}}{{\mu }}}-2\,\psi _{,\rho }}}}},  \label{energy} \\
L &=&\mu \,e^{-\psi }\,\sqrt{{\frac{{\psi _{,\rho }}}{{{\frac{{\mu _{,\rho}}%
}{{\mu }}}-2\,\psi _{,\rho }}}}},  \label{momentum} \\
\Omega &=&{\frac{{e^{2\psi }}}{{\mu }}}\,\sqrt{{\frac{{\psi _{,\rho }}}{{{%
\frac{{\mu _{,\rho }}}{{\mu }}}-\psi _{,\rho }}}}},  \label{angularv} \\
v^{(\varphi )} &=&\sqrt{{\frac{{\psi _{,\rho }}}{{{\frac{{\mu _{,\rho}}}{{\mu%
}}}-\psi _{,\rho }}}}},  \label{tangentialv}
\end{eqnarray}

\noindent and for the stability condition:

\begin{equation}
V_{,\rho \rho }|_{extr}=-{\frac{{2\,e^{2(\psi -\gamma )}}}{{{{\frac{{\mu
_{,\rho }}}{{\mu }}}-2\,\psi _{,\rho }}}}}\left( {\frac{{\mu _{,\rho }}}{{%
\mu }}}\,\psi _{,\rho \rho }-{\frac{{\mu _{,\rho \rho }}}{{\mu }}}\,\psi
_{,\rho }+4\,{\psi _{,\rho }}^{3}-6{\frac{{\mu _{,\rho }}}{{\mu }}}\,{\psi
_{,\rho }}^{2}+3\,\left( {\frac{{\mu _{,\rho }}}{{\mu }}}\right) ^{2}\,\psi
_{,\rho }\right) >0  \label{secderiv}
\end{equation}

\noindent where a coma stands for partial derivative.\newline

Now, observe that if the functions $\psi $ and $\mu $ are related by

\begin{equation}
e^{\psi }=({\frac{{\mu }}{{\mu _{0}}}})^{l}.  \label{main}
\end{equation}

\noindent being $l=const,$ we obtain a necessary and sufficient condition
for the velocity ${{v_{c}}^{(\varphi )}}$ to be the same for two orbits at
different radii, given $l={({v_{c}}^{(\varphi )})^{2}/}\left( {1+({v_{c}}%
^{(\varphi )})^{2}}\right) ,$ and equation (\ref{secderiv}) tells us that
this motion is stable. We call equation (\ref{main}) together with such
value of $l$ the {\it flat curve condition}.\newline

We now write the Einstein's equations $G_{\alpha \beta }=8\pi T_{\alpha
\beta }$ for an arbitrary energy momentum tensor for the line element (\ref
{metric}):

\begin{eqnarray}
{\mu }D^{2}\psi +D\mu \,D\psi  &=&4\pi \,\mu \,[e^{-2(\psi -\gamma
)}\,(e^{-2\psi }T_{tt}+{\frac{{e^{2\psi }}}{{\mu ^{2}}}}\,T_{{\varphi }{%
\varphi }})+T_{\rho \rho }+T_{zz}],  \label{ee1} \\
D^{2}\mu  &=&8\pi \,\mu \,[T_{\rho \rho }+T_{zz}]  \label{ee2} \\
\gamma _{\rho }\,\mu _{\rho }-\gamma _{z}\,\mu _{z}-\mu \,({\psi _{\rho }}%
^{2}-{\psi _{z}}^{2})+\mu _{zz} &=&8\,\pi \,\mu \,T_{\rho \rho },
\label{ee3} \\
\gamma _{\rho }\,\mu _{z}+\gamma _{z}\,\mu _{\rho }-2\,\mu \,\psi _{\rho
}\,\psi _{z}-\mu _{\rho z} &=&8\,\pi \,\mu \,T_{\rho z}.
\end{eqnarray}
where we have introduced the operator $D=(\partial _{\rho },\partial _{z})$,
see reference \cite{MNQ}. In order to have flat tangential curve velocities,
it is introduced the {\it flat curve condition} (\ref{main}). This condition
is valid on the equatorial plane. Nevertheless, the halo is expected to be
almost spherically symmetric, that means that if we know the functional
dependence of the gravitational potential on the equatorial plane, this
dependence should be the same one in almost the rest of the halo. In that
case it is reasonable to suppose that the {\it flat curve condition} (\ref
{main}) is valid in a region around the equatorial plane. Thus, in this
region we substitute the relations (\ref{main}) into the left hand side of
equation (\ref{ee1}) obtaining $\mu \,D^{2}\psi +D\mu \,D\psi =lD^{2}\mu $
and with (\ref{ee2})\ we get a constrain equation amount the components of
the stress energy tensor:

\begin{equation}
-\left( {\frac{{1-({v_{c}}^{(\varphi )})^{2}}}{{1+({v_{c}}^{(\varphi )})^{2}}%
}}\right) (T_{\rho \rho }+T_{zz})=e^{-2(\psi -\gamma )}\left( e^{-2\psi
}T_{tt}+{\frac{{e^{2\psi }}}{{\mu ^{2}}}}\,T_{{\varphi }{\varphi }}\right) 
\label{ee33}
\end{equation}

Notice that this relation must be satisfied by any stress energy tensor
which, within the approximation made in the analysis, curves the space
time in such a way that the motion of test particles corresponds to the
observed one.\\

Let us consider the case of a stress energy tensor corresponding to a
perfect fluid, $T_{\mu \nu }=(d+p)\,u_{\mu }\,u_{\nu }+g_{\mu \nu }\,p$,
with $d$ the density of the fluid and $p$ its pressure. In this case we are
thinking on a ``dark fluid'', which is not seen but it is thought that it
could be there affecting the geometry in the way needed in order to have the
observed behavior in the tangential velocities of the luminous matter, as
just mentioned. Considering the dark fluid as static, the four velocity of
such dark fluid is given by $u^{\alpha }=(u^{0},0,0,0)$ which, for the line
element (\ref{metric}) reads: $u^{0}=E\,e^{-2\psi }$, thus $u_{0}=-E$ and
from $u^{\mu }u_{\mu }=-1$, we obtain that $E=e^{2\psi }$. Therefore, the
stress energy tensor has the form:

\begin{eqnarray}
&&T_{tt}=e^{2\psi }d,  \label{T00} \\
&&T_{\rho \rho }=T_{zz}=e^{-2(\psi -\gamma )}p,  \label{T11} \\
&&T_{{\varphi }{\varphi }}=\mu ^{2}\,e^{-2\psi }p.  \label{T33}
\end{eqnarray}

Substituting these expressions into (\ref{ee33}), we obtain that in the
equatorial plane, in order to satisfy the observed behavior on the
tangential velocities, the ``dark fluid'' has to fulfill the relation:

\begin{equation}
-2\left( {\frac{{1-({v_{c}}^{(\varphi )})^{2}}}{{1+({v_{c}}^{(\varphi )})^{2}%
}}}\right) \,p=(d+p)\,  \label{sesemifinal}
\end{equation}

Let us see which are the permitted relations between pressure and density of
the perfect fluid providing flat rotation curves, we thus obtain:

\begin{equation}
\,p=-{\frac{{1+({v_{c}}^{(\varphi )})^{2}}}{{3-({v_{c}}^{(\varphi )})^{2}}}d}
\label{sefinal}
\end{equation}

\noindent relation from which the $d$ coefficient is identified with the
square velocity dispersion of the dark particles, that appears to be
negative. We are now in a convenient position to strict the state equation.
As the velocities of the gas and stars rotating in the flat region must be
within $0<v_{c}^{(\varphi )2}<1$, (the observed ones are of the order of $%
v_{c}^{(\varphi )}\sim 10^{-3}$ \cite{persic}), relation (\ref{sefinal})
implies $-1<\omega <-1/3$, being $p=\omega d$. This result
coincides with the one obtained at cosmological scale for the respective
equation of state in the Quintessence model \cite{stein,DM-CQG}.
\newline

Therefore the analysis presented in this letter, gives support to the
hypothesis that a Quintessence-like equation of state could be the solution
for the dark matter problem at galactic scale. In both cases it turns out
the need of an exotic equation of state, with $\omega =-0.33$ at a galactic
scale and $\omega =-0.64$ for the cosmos \cite{luis}.\\

In any case, we have shown that galactic dark matter satisfying an exotic
equation of state certainly can be used to explain the observed behavior
on the rotational curves of spiral galaxies.\newline

We want to thank Daniel Sudarsky, and Alejandro Corichi for many helpful
discussions. We also want to express our acknowledgment to the relativity
group in Jena for its kind hospitality and partial support. This work is
also partly supported by CONACyT M\'{e}xico, 94890 (Guzm\'{a}n) and by the
DGAPA-UNAM IN121298 (N\'{u}\~{n}ez, Ram\'{i}rez).\newline



\begin{references}
\bibitem{riess} S.  Perlmutter {\it et al}. {\it ApJ,} {\bf 517} (1999)
565. A. G. Riess {\it et al.}, {\it Astron.J}, {\bf 116} (1998) 1009-1038.

\bibitem{luis}  T. Matos, F. S. Guzm\'an and L. A. Ure\~na, 
{\it Class. Quant. Grav.}, {\bf 17} (2000) 1707-1712.

\bibitem{stein}  R. R. Caldwell, Rahul Dave and Paul J. Steinhardt, {\it %
Phys.Rev. Lett.}, {\bf 80} (1988) 1582-1585. Ivaylo Zlatev, Limin Wang and
Paul J. Steinhardt, {\it Phys.Rev. Lett.}, {\bf 82} (1988) 896-899.

\bibitem{rubin} V. C. Rubin, N. Thonnard, W. K. Ford and M. S. Roberts,
{\it ApJ} {\bf 81} (1997) 719.

\bibitem{apj}  Philip D. Mannheim, {\it ApJ} {\bf 479} (1997) 659.

\bibitem{persic}  Massimo Persic, Paolo Salucci and Fulvio Stel, 
{\it MNRAS}, {\bf 281} (1996) 27-47.

\bibitem{peebles}  P. J. E. Peebles, {\it Principles of Physical Cosmology},
Princeton University Press, Princeton, (1993).

\bibitem{DM-CQG}  F. Siddhartha Guzm\'{a}n and Tonatiuh Matos, {\it Class.
Quant. Grav.,} {\bf 17} (2000) L9-L16. F. S. Guzm\'{a}n, T. Matos and H.
Villegas. Preprint astro-ph/9811143.

\bibitem{Nvo}  T. Matos, D. N\'{u}\~{n}ez, F. S. Guzm\'{a}n and E.
Ram\'{i}rez, {\it Geometric conditions on the type of matter determining the
flat behavior of the rotational curves in galaxies,} in preparation.

\bibitem{chandra} S. Chandrasekhar, {\it Mathematical theory of black
holes}, Oxford Science Publications (1983).

\bibitem{MNQ}  T. Matos, D. N{\'{u}}{\~{n}}ez, H. Quevedo {\it Phys. Rev}, 
{\bf D 51} (1995) R310-R313.
\end{references}
\end{document}